\documentclass[onecolumn,11pt,superscriptaddress]{revtex4}
\usepackage{amsmath,amssymb}
\usepackage{bm}
\usepackage{graphicx}
\usepackage{hyperref}
\usepackage[dvips]{color}
\usepackage[]{natbib}
\begin{document}

\title{A Novel Method to Determine Magnetic Fields in low-density Plasma e.g. Solar Flares
Facilitated Through Accidental Degeneracy of Quantum States in Fe$^{9+}$}

\author{Wenxian Li}
\affiliation{The Key lab of Applied Ion Beam Physics, Ministry of Education, China}
\affiliation{Shanghai EBIT laboratory, Institute of Modern physics, Fudan University, Shanghai, China}

\author{Jon Grumer}
\affiliation{Division of Mathematical Physics, Department of Physics, Lund University, Sweden}

\author{Yang Yang}
\affiliation{The Key lab of Applied Ion Beam Physics, Ministry of Education, China}
\affiliation{Shanghai EBIT laboratory, Institute of Modern physics, Fudan University, Shanghai, China}

\author{Tomas Brage}\email{Tomas.Brage@fysik.lu.se}
\affiliation{Division of Mathematical Physics, Department of Physics, Lund University, Sweden}

\author{Ke Yao}
\affiliation{The Key lab of Applied Ion Beam Physics, Ministry of Education, China}
\affiliation{Shanghai EBIT laboratory, Institute of Modern physics, Fudan University, Shanghai, China}

\author{Chongyang Chen}
\affiliation{The Key lab of Applied Ion Beam Physics, Ministry of Education, China}
\affiliation{Shanghai EBIT laboratory, Institute of Modern physics, Fudan University, Shanghai, China}

\author{Tetsuya Watanabe}
\affiliation{ National Astronomical Observatory of Japan (NAOJ), Tokyo, Japan}

\author{Per J\"onsson}
\affiliation{Division of Material Science and Computational Mathematics, Malm\"o University, Sweden}

\author{Henrik Lundstedt}
\affiliation{Swedish Institute of Space Physics, Solar-Terrestrial Physics Division, Lund, Sweden}

\author{Roger Hutton}\email{rhutton@fudan.edu.cn}
\affiliation{The Key lab of Applied Ion Beam Physics, Ministry of Education, China}
\affiliation{Shanghai EBIT laboratory, Institute of Modern physics, Fudan University, Shanghai, China}

\author{Yaming Zou}
\affiliation{The Key lab of Applied Ion Beam Physics, Ministry of Education, China}
\affiliation{Shanghai EBIT laboratory, Institute of Modern physics, Fudan University, Shanghai, China}

\begin{abstract}
We propose a new method to determine magnetic fields, by using the magnetic-field induced electric dipole transition $3p^4\,3d~^4\mathrm{D}_{7/2}$  $\rightarrow$  $3p^5~^2\mathrm{P}_{3/2}$ in Fe$^{9+}$ ions. This ion has a high abundance in astrophysical plasma and is therefore well-suited for direct
measurements of even rather weak fields in e.g. solar flares. This transition is induced by an external
magnetic field and its rate is proportional to the square of the magnetic field
strength. We present theoretical values for what we will label the reduced
rate and propose that the critical energy difference between the upper level in this transition and the close to degenerate $3p^4\,3d~^4\mathrm{D}_{5/2}$ should be measured experimentally since it is required to determine the relative intensity of this magnetic line for different magnetic fields.
\end{abstract}

\date{\today}

\maketitle

\section{Introduction}
One of the underlying causes behind solar events, such as solar flares, is the conversion of magnetic to thermal energy. It is therefore vital to be able to measure the magnetic field of the corona, over hot active areas of the sun which exhibits relatively strong magnetic fields.
In order to follow the evolution of a solar flare, continuous observations are required, either from space or by using a network of ground-based instruments. It is therefore unfortunate that there are no space-based coronal magnetic field measurements, but only model estimates based on extrapolations from measurements of the photospheric fields ({Schrijver et al. 2008}). Ground-based measurements are performed either in the radio range ({White 2004}) across the solar corona, or in the infrared wavelength range ({Lin et al. 2004}) on the solar limb. Infrared measurements of magnetic fields are limited by the fact that the spectral lines under investigation are optically thin. On the other hand, gyroresonance emission is optically thick, but refers only to a specific portion of the corona, which has a depth of around 100 km. From these measurements an absolute field strength at the base of the corona, above active regions, in the range of 0.02 - 0.2 T was obtained ({White 1997}).

In this work we present a completely new method to measure magnetic fields of the active corona. This method is based on an exotic category of light generation, fed by the plasma magnetic field, external to the ions, in contrast to the internal fields generated by the bound electrons. The procedure relies on radiation in the soft x-ray region of the spectrum, implying a space based method. This ¡°magnetic-field induced¡± radiation originates from atomic transitions where the lifetime of the upper energy level is sensitive to the local, external magnetic field ({Beiersdorfer et al. 2003, Li et al. 2013, Grumer et al. 2013, Grumer et al. 2014, Li et al. 2014}). We will show that there is a unique case where even relatively small external magnetic fields can have a striking effect on the ion, leading to resonant magnetic-field induced light, due to what is called accidental degeneracy of quantum states.

The impact of the coronal magnetic field on the ion is usually very small due to the relative weakness of these fields in comparison to the strong internal fields of the ions. The effect therefore usually only contributes very weak lines that are impossible to observe. However, sometimes the quantum states end up very close to each other in energy, they are accidentally degenerate, and the perturbation by the external field will be enhanced. If this occurs with a state that without the field has no, or only very weak, electromagnetic transitions to a lower state, a new and distinct feature in the spectrum from the ion will appear $-$ a new strong line. Unfortunately, since the magnetic fields internal to the ion and externally generated in the coronal plasma differs by about five to seven orders of magnitude, the probability of a close-enough degeneracy is small. But in this report we will discuss a striking case of accidental degeneracy in an important ion for studies of the sun and other stars, Fe$^{9+}$.

The origin of the new lines in the spectra of ions is the breaking of the atomic symmetry by the external field, which will mix atomic states with the same magnetic quantum number and parity. This will in turn introduce new decay channels from excited states ({Andrew et al. 1967, Wood et al. 1968}), which we will label magnetic-field induced transitions (MITs) ({Grumer et al. 2014}). These transitions have attracted attention recently, when accurate and systematic methods to calculate their rates have been developed ({Grumer et al. 2013, Li et al. 2013}).

\begin{figure}
\centering
\includegraphics[width=0.6\textwidth]{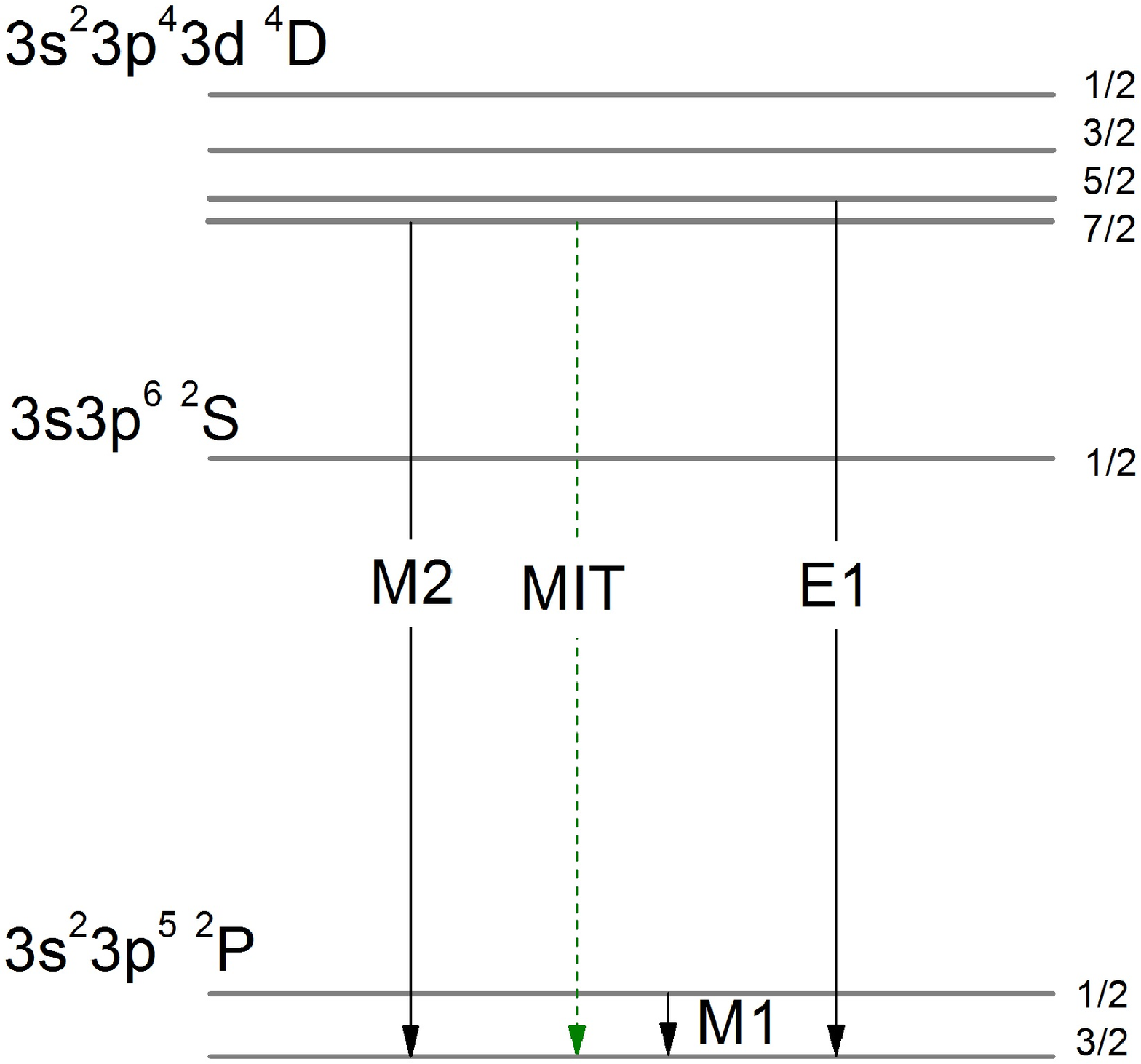}
\caption{\label{level}(Color online) Schematic energy-level diagram for Chlorine-like ions with $Z~\textless~26 $ and zero nuclear spin, where $^4\mathrm{D}_{7/2}$ is the lowest level in the configuration $3s^23p^43d$. For ions with $ Z~\textgreater~26$, a level crossing has
occured and $^4\mathrm{D}_{5/2}$ is lower than $^4\mathrm{D}_{7/2}$. Under the influence of an external magnetic field, an E1 transition opens up from the $^4\mathrm{D}_{7/2}$ to the ground state through mixing with the $^4\mathrm{D}_{5/2}$.}
\end{figure}

\section{Structure of Chlorine-like ions and MITs}
\begin{figure}[h]
\centering
\includegraphics[width=0.8\textwidth]{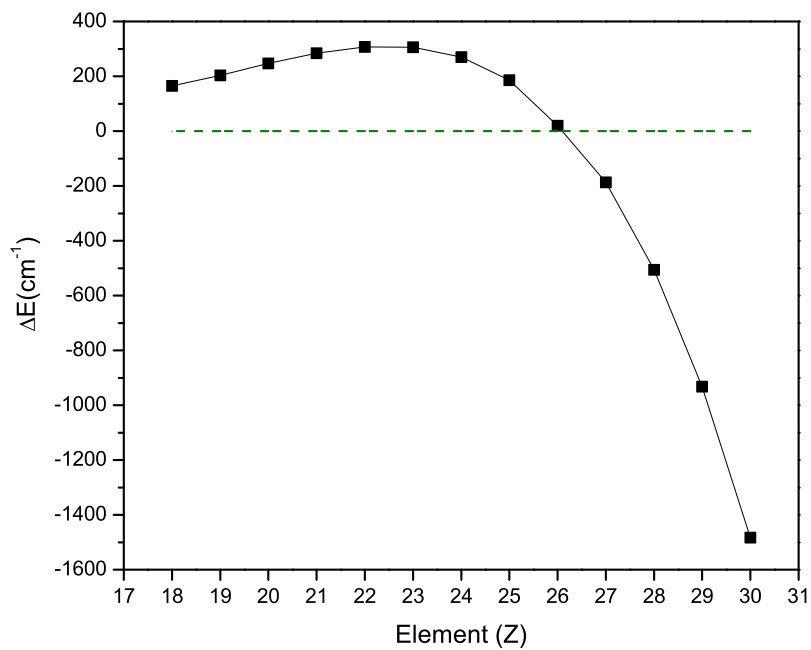}
\caption{\label{dEsequence} The fine-structure splitting energy $\Delta E = E(^4\mathrm{D}_{7/2}) - E(^4\mathrm{D}_{5/2})$ as a function of the nuclear charge, along the isoelectronic sequence from calculations reported here. The dashed line in green marks $\Delta E = 0$.}
\end{figure}

The structure of the lowest levels of Chlorine-like ions is illustrated in Figure \ref{level}. The important levels in the present study are the two lowest in the term $3p^43d~^4\mathrm{D}^e$, which turn out to have very different decay modes. Without external fields and ignoring the effects of the nuclear spin, they both decay to the $3p^5~^2\mathrm{P}^o_{3/2}$-level in the ground configuration, but while the $J=5/2$ has
a fast electric dipole (E1) channel, the $J=7/2$ can only decay via a slow magnetic quadrupole (M2) transition. In the presence of an external magnetic
field, these two states will mix and induce an E1, MIT competing transition channel from the $J=7/2$ level. For most ions the M2 transition is still the dominant decay channel, but a crossing of the fine structure levels $^4\mathrm{D}_{7/2}$ and $^4\mathrm{D}_{5/2}$ between Cobalt and Iron (see Figure. \ref{dEsequence}) will change the picture. As a matter of fact, for Iron this fine-structure splitting energy is predicted to be at a minimum and the MIT-contribution to the decay of the  $J=7/2$ level will be strongly enhanced.

Unfortunately, there are large variations in the predicted value of this energy difference (see Table~\ref{iron}). The aim of this report is therefore to (a)  make a careful theoretical study of the energy splitting between the two $^4\mathrm{D}^e$-levels along the isoelectronic sequence, to confirm the close degeneracy for iron, and (b) to make an accurate prediction of what we will label the reduced decay rate, $a^R_{MIT}$ (see next section). This reduced rate can be combined by the experimentally determined wavelength and energy splitting to give the MIT-rate for different magnetic fields.

\begin{table}
\caption{\label{iron} \footnotesize ATOMIC DATA FOR FE X}
\vspace{0.1cm}
\begin{tabular}{llllllllllllllllll}
\hline
\hline
&&  Method                    &&   &&  $\lambda$        &&   &&  $\Delta E$   &&   &&        $A_{E1}$  \\
\hline
Observation &&Solar ({Thomas et al. 1994, Brosius et al. 1998})&&   &&     257.25        &&   &&   0           &&   &&                  \\
    &&Solar ({Sandlin 1979})         &&   &&                   &&   &&   5           &&   &&                  \\
\hline
  &&  present                          &&   &&    257.7285       &&   &&   20.14        &&   &&         6.30[6]  \\
  &&  MCDF ({Huang et al. 1983})              &&   &&     246.4924      &&   &&   78          &&   &&         1.63[6]  \\
  &&  MCDF ({Dong et al. 1999})              &&   &&     256.674       &&   &&   108         &&   &&         6.27[6]  \\
Theory  &&  MCDF ({Aggarwal et al. 2004})            &&   &&                   &&   &&   54.85       &&   &&                  \\
  &&  MR-MBPT ({Yasuyuki et al. 2010})           &&   &&     257.1924      &&   &&   18          &&   &&                  \\
  &&  R-matrix ({Del Zanna et al. 2012})             &&   &&     246.8890      &&   &&   109.74      &&   &&                  \\
  &&  CI ({Bhatia et al. 1995})              &&   &&     256.1974      &&   &&   $-$58       &&   &&         1.21[6]  \\
  &&  CI ({Deb et al. 2002})                &&   &&     257.0846      &&   &&   21          &&   &&         2.42[5]  \\
\hline
\hline
\end{tabular}
\footnotetext{NOTES. --List of the wavelength $\lambda$ (in \AA) of the magnetically induced $3p^43d~^4\mathrm{D}_{7/2} \rightarrow 3p^5~^2\mathrm{P}_{3/2}$ transition, the fine structure splitting $\Delta E = E\left(^4\mathrm{D}_{5/2}\right)-E\left(^4\mathrm{D}_{7/2}\right)$ (in cm$^{-1}$) and the
rate $A_{E1} (3p^43d~^4\mathrm{D}_{5/2}\rightarrow 3p^5~^2\mathrm{P}_{3/2})$ of the E1 transition (in s$^{-1}$) for Chlorine-like Iron. The second column lists the various sources. Numbers in square brackets are the power of 10.}
\end{table}

\section{Theoretical method and Computational model}
\subsection{Theoretical method}
The basis of our theoretical approach is described in our earlier papers on MITs ({Li et al. 2013, Grumer et al. 2013}).
In our example the reference state is $|3p^43d~^4\mathrm{D}_{7/2}\rangle$, which we can
represent by a mixture of two pure states in the presence of a magnetic field
\begin{small}
\begin{eqnarray}
\label{WF-1}
|``3p^43d~^4\mathrm{D}_{7/2}" ~ M  \rangle = d_0 | 3p^43d~^4\mathrm{D}_{7/2} ~ M \rangle + d_1(M)|3p^43d~^4\mathrm{D}_{5/2} ~ M \rangle.
\end{eqnarray}
\end{small}
where we ignore interactions with other atomic states, since their energies are far from the reference state.
The total E1 transition rate from a specific $M$ sublevel in the mixed $``3p^43d~^4\mathrm{D}_{7/2}"$ to all the $M'$ sublevels of the ground level $3p^5~^2\mathrm{P}_{3/2}$ can be expressed as:
\begin{small}
\begin{eqnarray}
\label{MIT-2}
A_{MIT}(M) = \sum_{M'} A_{MIT}(M,M') \approx \frac{2.02613 \times 10^{18}} {3 \lambda^3} \left |\ d_1(M) \langle 3p^5~^2\mathrm{P}_{3/2} || {\bf P}^{(1)} || 3p^43d~^4\mathrm{D}_{5/2} \rangle \right |^2.
\end{eqnarray}
\end{small}
where $\lambda$ is the transition wavelength and $d_1(M)$ depends on the magnetic quantum number $M$ of the sublevels belonging to the $3p^43d~^4\mathrm{D}_{7/2}$ level. For the $3p^43d~^4\mathrm{D}_{7/2}$ level, $d_1(M)$ is given by
\begin{small}
\begin{eqnarray}
\label{MIT-3}
d_1(M) &=& \frac{\langle ~^4\mathrm{D}_{5/2} M | H_m | ^4\mathrm{D}_{7/2} M \rangle}{E(^4\mathrm{D}_{7/2}) - E(^4\mathrm{D}_{5/2})} \nonumber \\
                     &=& -B \sqrt{\frac{49-4 M^2}{63}} \frac{\langle ^4\mathrm{D}_{5/2} || {\bf N}^{(1)} + \Delta {\bf N}^{(1)} || ^4\mathrm{D}_{7/2} \rangle}{E(^4\mathrm{D}_{7/2}) - E(^4\mathrm{D}_{5/2})}.
\end{eqnarray}
\end{small}
As a result, the total rates of the $3p^43d~^4\mathrm{D}_{7/2}\rightarrow 3p^5~^2\mathrm{P}_{3/2}$ MITs from individual sublevels can be expressed as
\begin{small}
\begin{eqnarray}
\label{MIT4}
A_{MIT}(M) &=&a^R_{MIT}(M)\frac{B^2}{\lambda^3(\Delta E)^2}.
\end{eqnarray}
\end{small}
where $B$ is in units of T, $\lambda$ is in units of \AA, $\Delta E = E(^4\mathrm{D}_{7/2}) - E(^4\mathrm{D}_{5/2})$(in units of cm$^{-1}$) and we have defined a reduced transition rate as
\begin{small}
\begin{equation}
\label{MIT5}
a^R_{MIT}(M) \approx \frac{2.02613 \times 10^{18}\cdot(49-4 M^2)}{189}
\left |\langle ^4\mathrm{D}_{5/2} || {\bf N}^{(1)} + \Delta {\bf N}^{(1)} || ~^4\mathrm{D}_{7/2} \rangle \langle ^3P_{3/2} || {\bf P}^{(1)} || ^4\mathrm{D}_{5/2} \rangle \right|^2 .\end{equation}
\end{small}
The reduced rate defined in this equation is independent of the transition wavelength, the magnetic field strength as well as the energy splitting. This gives us the property that relates the MIT-rates to the external magnetic field strength. To determine $A_{MIT}(M)$, we recommend to use theoretical values of $a_{MIT}^R(M)$, as reported in this work, combined with experimental values of the energy splitting and wavelength.

\subsection{Correlation Model}
The calculations are based on the Multiconfiguration Dirac-Hartree-Fock (MCDHF) method, in the form of the latest version of the GRASP2K program ({J\"onsson et al. 2013}). A single reference configuration model is adopted for the even-parity($3p^43d$) and odd-parity($3p^5$) states, and the $1s$, $2s$, $2p$ core subshells are kept closed. The set of CSFs is obtained by single and double excitations from the n=3 shell of the reference configurations to the active set. The active set is augmented layer by layer to n=7 ($l_{max}=4$) when satisfactory convergence is achieved. For each step, we optimize only the orbitals in the last added correlation layer at the time. In the final calculations, the total number of CSFs was 16490 for the odd-parity(J=3/2, 1/2) and 523421 for the even-parity (J = 1/2, 3/2, 5/2, 7/2, 9/2) cases.

The resulting excitation energies of the  $^4\mathrm{D}_{7/2}$ and $^4\mathrm{D}_{5/2}$ changes by less than 0.1\% in the last step of the calculation. The final excitation energies agree with experiment to within 1\%. The crucial energy splitting between $^4\mathrm{D}_{7/2}$ and $^4\mathrm{D}_{5/2}$ is well-converged, except for iron where the close degeneracy occurs. To better represent this critical case we extended the calculations to include single excitations from the $2s$ and $2p$ subshells.

\section{Results and Discussion}
\subsection{Isoelectronic Sequence}
We present in Table~\ref{AEH} all the important properties, according to Eq. (\ref{MIT4}), involved in computing the MIT-rates, i.e. the reduced transition rate $a^R_{MIT}(M)$, the energy splitting, $\Delta E$, between the two levels $3p^43d~^4\mathrm{D}_{5/2}$ and $^4\mathrm{D}_{7/2}$, together with the wavelength, $\lambda$, of the $3p^43d~^4\mathrm{D}_{7/2} \rightarrow 3p^5~^2\mathrm{P}_{3/2}$ transition.

\begin{table}
\caption{\label{AEH} \footnotesize CALCULATIONAL RESULTS FOR CL-LIKE IONS}
\vspace{0.1cm}
\begin{tabular}{ccccccccccccccccccccccccc}
\hline
\hline
  &&  && \multicolumn{5}{c}{$a^R_{MIT}(M)$} &&         &&          \\
\cline{4-9} ions   && $A_{M2}$ &&  $M=\pm {1/2}$ && $M=\pm {3/2}$   &&   $M=\pm {5/2}$ && $\Delta E$ && $\lambda$ \\

\hline
Ar$^{+}$     &&   1.26[0]    &&     7.994[0]  &&	6.662[0]  &&	3.997[0]       &&     165.5         &&       762.5978    \\
 K$^{2+}$    &&   3.20[0]    &&     7.686[0]  &&	6.405[0]  &&	3.843[0]       &&     202.87        &&       600.4231    \\
Ca$^{3+}$    &&   6.16[0]    &&     7.621[0]  &&	6.351[0]  &&	3.810[0]       &&     246.65        &&       500.1306    \\
Sc$^{4+}$    &&   1.03[1]    &&     7.910[0]  &&	6.592[0]  &&	3.955[0]       &&     283.9         &&       430.6129    \\
Ti$^{5+}$    &&   1.59[1]    &&     8.443[0]  &&	7.036[0]  &&	4.222[0]       &&     307.02        &&       379.0310    \\
 V$^{6+}$    &&   2.31[1]    &&     9.093[0]  &&	7.577[0]  &&	4.546[0]       &&     306.35        &&       338.9583    \\
Cr$^{7+}$    &&   3.22[1]    &&     9.807[0]  &&	8.172[0]  &&	4.903[0]       &&     270.41        &&       306.7765    \\
Mn$^{8+}$    &&   4.33[1]    &&     1.055[1]  &&	8.793[0]  &&	5.276[0]       &&     186.08        &&       280.2679    \\
Fe$^{9+}$    &&   5.68[1]    &&     1.119[1]  &&	9.326[0]  &&	5.594[0]       &&     20.14         &&       257.7285    \\
Co$^{10+}$   &&   7.30[1]    &&     1.201[1]  &&	1.001[1]  &&	6.006[0]       &&     $-$186.87     &&       238.9746    \\
Ni$^{11+}$   &&   9.20[1]    &&     1.261[1]  &&	1.051[1]  &&	6.305[0]       &&     $-$505.53     &&       222.5113    \\
Cu$^{12+}$   &&   1.14[2]    &&     1.309[1]  &&	1.090[1]  &&	6.543[0]       &&     $-$932.75     &&       208.1047    \\
Zn$^{13+}$   &&   1.40[2]    &&     1.347[1]  &&	1.123[1]  &&	6.736[0]       &&     $-$1482.74    &&       195.3790    \\
\hline
\hline
\end{tabular}
\footnotetext{NOTES. --List of the M2 transition rates, $A_{M2}$ (in s$^{-1}$), the reduced transition rate $a^R_{MIT}(M)$ (in $ 10^{12}$ $\rm{\AA^3 \cdot cm^{-2} \cdot T^{-2}\cdot s^{-1}}$), and the wavelength $\lambda$ (in \AA) of the MIT-transition $3p^43d~^4\mathrm{D}_{7/2}\rightarrow 3p^5~^2\mathrm{P}_{3/2}$ together with the energy splitting $\Delta E  = E(^4\mathrm{D}_{5/2})-E(^4\mathrm{D}_{7/2})$ (in cm$^{-1}$) for Cl-like Ar to Zn.  x[n] indicates $x\times 10^n$.}
\end{table}

In the absence of an external magnetic field, magnetic quadrupole (M2) is the dominant decay channel for the $3p^43d~^4\mathrm{D}_{7/2}\rightarrow 3p^5~^2\mathrm{P}_{3/2}$ transition. When an external magnetic field is introduced, an additional decay channel is opened and we define a average transition rate $\overline{A}_{MIT}$ of the $3p^43d~^4\mathrm{D}_{7/2} \rightarrow 3p^5~^2\mathrm{P}_{3/2}$ transition,
\begin{small}
\begin{equation}
\label{tau3}
\overline{A}_{MIT} = \frac{\sum_M A_{MIT}(M)}{2J+1}.
\end{equation}
\end{small}

\begin{figure}
\centering
\includegraphics[width=0.8\textwidth]{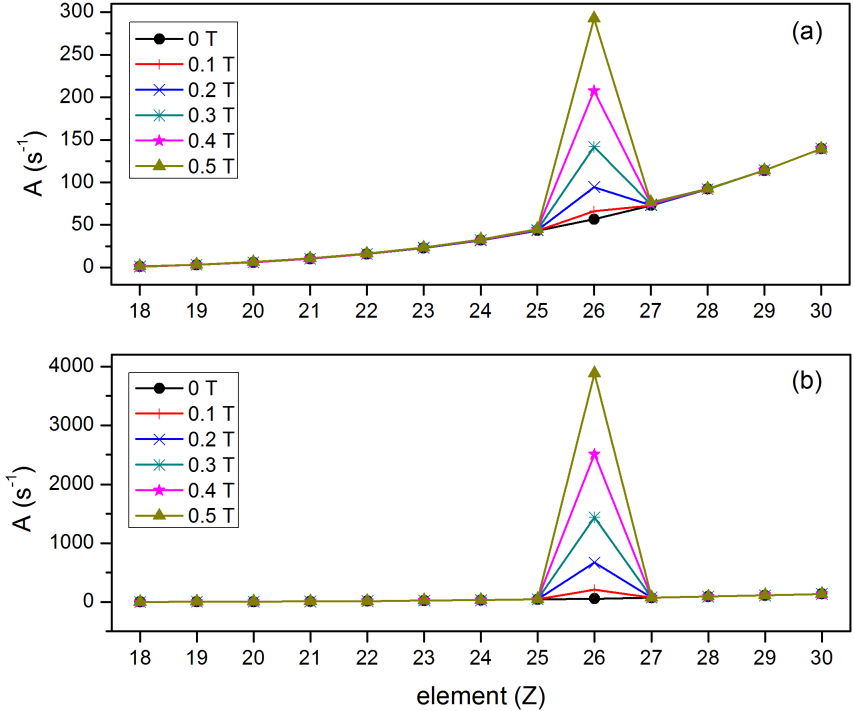}
\caption{\label{averagetr}
(Color online) The total transition rate $A~=~A_{M2}~+~\overline{A}_{MIT}$ of the $3p^43d~^4\mathrm{D}_{7/2} \rightarrow 3p^5~^2\mathrm{P}_{3/2}$ transition along the Cl-like isoelectronic sequence for some magnetic-field strengths. We used the fine structure energy of (a)~20.14 ~$\mathrm{cm}^{-1}$ and (b)~5 ~$\mathrm{cm}^{-1}$ for iron, respectively.}
\end{figure}

Rates for any field can be obtained by eq.~\ref{MIT4} and \ref{MIT5}. We plot the transition rates $A~=~A_{M2}~+~\overline{A}_{MIT}$ along the isoelectrionic sequence in Figure~\ref{averagetr}~(a) for some magnetic-field strengths and $\Delta E~=~20.14~\mathrm{cm}^{-1}$ for Iron. It is clear that the magnetic field influences the transition rate substantially for Iron due to the close degeneracy. To further illustrate the resonance behaviour of this effect, we also used an astrophysical value~({Sandlin 1979})~$\Delta E~=~5~\mathrm{cm}^{-1}$ for Iron, in Figure~\ref{averagetr}~(b).

\subsection{Fe X}
Due to the close to complete cancellation for iron of the energy difference between the two
$^4\mathrm{D}$-levels (see Figure~\ref{dEsequence}) we will pay special attention to this
ion.

It is clear that some of the properties in Table~\ref{AEH} are more easily obtainable through theoretical calculations. We illustrate this in Table~\ref{WS} where we
show the convergence of the calculated off-diagonal reduced matrix elements, W = $\langle ^4\mathrm{D}_{5/2} || {\bf N}^{(1)} + \Delta {\bf N}^{(1)} || ~^4\mathrm{D}_{7/2} \rangle$, representing the magnetic interaction, together with the line strength, S = $\left |\langle ^3P_{3/2} || {\bf P}^{(1)} || ^4\mathrm{D}_{5/2} \rangle \right|^2$, of the close-lying E1 transition. Since these values converges fast and are not subjected to cancellation effects, we estimate their
accuracy to be well within a few percent. This will in turn imply that the prediction of the reduced transition rate $a^R_{MIT}(M)$ in Equation~\ref{MIT4} and \ref{MIT5}
is of similar accuracy.
 \begin{table}
\caption{\label{WS} \footnotesize CONVERGENCE STUDY OF THE CALCULATIONS}
\vspace{0.1cm}
\begin{tabular}{ccccccccccccccccccccccccc}
\hline
\hline
layer  && && W  && && S &&  \\
\hline
DF      &&  &&   0.5305 &&  &&   4.063[4]      \\
n=4     &&  &&   0.5285 && 	&&   3.555[4]      \\
n=5     &&  &&   0.5284 && 	&&   3.355[4]      \\
n=6     &&  &&   0.5283 && 	&&   3.298[4]      \\
n=7     &&  &&   0.5283 && 	&&   3.264[4]      \\
\hline
\hline
\end{tabular}
\footnotetext{NOTES. --Convergence study of calculated off-diagonal reduced matrix elements W = $\langle ^4\mathrm{D}_{5/2} || {\bf N}^{(1)} + \Delta {\bf N}^{(1)} || ~^4\mathrm{D}_{7/2} \rangle$ (in a.u.) of the magnetic interaction and line strength S = $\left |\langle ^3P_{3/2} || {\bf P}^{(1)} || ^4\mathrm{D}_{5/2} \rangle \right|^2$ (in a.u.). DF represents a single configuration Dirac-Hartree-Fock model. x[n] indicates $x\times 10^n$.}
\end{table}

\begin{figure}[h]
\centering
\includegraphics[width=0.8\textwidth]{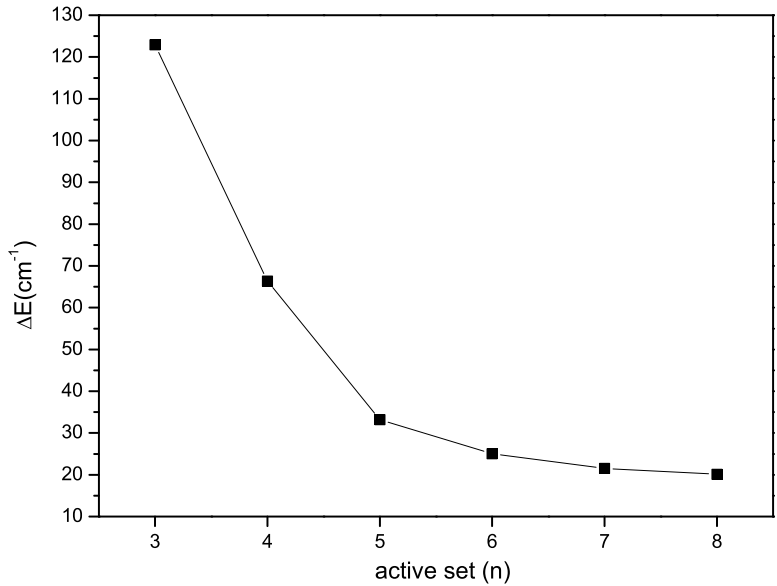}
\caption{\label{deltaE} Convergence trend of the fine-structure separation $\Delta E= E(^4\mathrm{D}_{5/2}) -E(^4\mathrm{D}_{7/2})$ for Fe$^{9+}$ as the size of the active set of orbitals is increased as defined by the maximum n-quantum number.}
\end{figure}

 We give in Figure~\ref{deltaE} the energy difference $\Delta E$ as a function of the maximum $n$ in the active set and thereby of the size of the CSF-expansion. It is clear that we also here reach a convergence close to a few cm$^{-1}$ for this property. The final value for the fine structure splitting is 20.14 cm$^{-1}$, in good agreement with the result from recent configuration interaction calculations ({Deb et al. 2002}) as well as Many-Body Perturbation Theory ({Ishikawa et al. 2010}). This strongly supports the prediction of the close degeneracy of the two levels for iron.

The strong resonance effect for iron is especially fortunate due to its high abundance in many astrophysical plasma. As a matter of fact, the ground transition is one of the ``coronal" lines used to determine the temperature of the corona and is known as the corona red line ({Swings 1943}). Unfortunately there is no firm experimental value for the critical energy splitting between the $^4\mathrm{D}_{7/2}$ and $^4\mathrm{D}_{5/2}$. At the same time, it is a great challenge to calculate the size of this accidental degeneracy accurately.

\begin{figure}[t]
\centering
\includegraphics[width=0.8\textwidth]{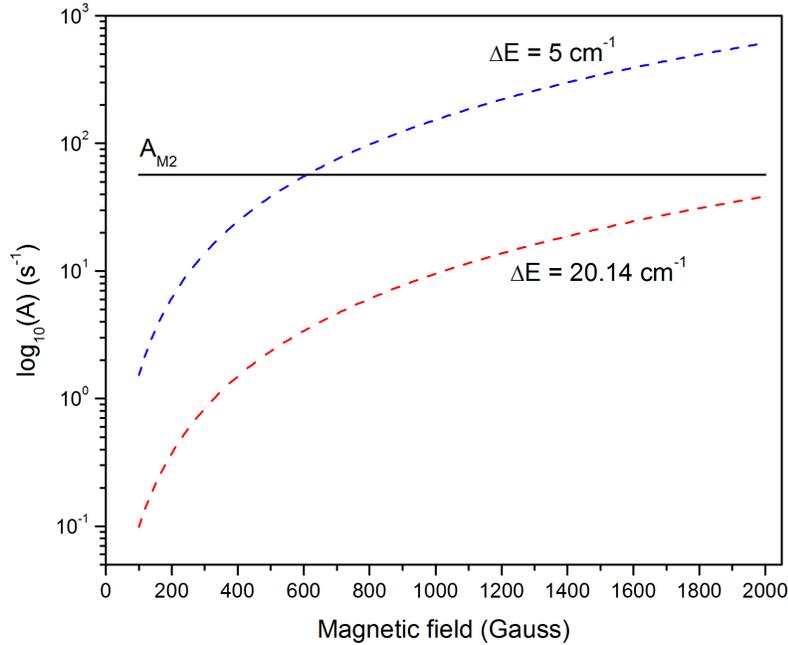}
\caption{\label{Btr} A plot of the $\overline{A}_{MIT}$ as a function of magnetic fields at $\Delta E = 20.14$ cm$^{-1}$ and $\Delta E = 5$ cm$^{-1}$, and compared to $A_{M2}$. }
\end{figure}

In the early experimental work by Smitt ({Smitt 1977}) the two levels were given identical excitation energies of 388708 cm$^{-1}$. Since then there has been a great deal of work by different groups to study the structure of Fe X (see Table \ref{iron}). The differences between the various calculations are often much larger than the predicted fine-structure splitting, which leads to large uncertainties in the level ordering and line identifications. Huang ({Huang et al. 1983}) and Dong et al ({Dong et al. 1999}) performed multi-configuration Dirac-Fock (MCDF) calculation and predicted the $^4\mathrm{D}_{5/2,7/2}$ levels to be separated by 78 cm$^{-1}$ and 108 cm$^{-1}$ respectively. Ishikawa ({Ishikawa et al. 2010}) predicted 18 cm$^{-1}$ from his Multireference-MBPT method. There is a recommended value from solar observations of around 5 cm$^{-1}$, determined from short-wavelength transitions from higher levels and therefore probably quite uncertain. Predictions from The Goddard Solar Extreme Ultraviolet Rocket Telescope and Spectrograph SERTS$-$89 ({Thomas et al. 1994}) and SERTS$-$95 ({Brosius et al. 1998}) spectra give the same energy for $^4\mathrm{D}_{7/2}$ and $^4\mathrm{D}_{5/2}$, probably due to limited resolution. Finally Del Zanna ({Del Zanna et al. 2004}) benchmarked the atomic data for Fe X and suggested the best splitting energy to be 5 cm$^{-1}$.
Although our calculations reach a convergence within the model, to the final value of around 20 cm$^{-1}$, it is clear that systematic errors, such as omitted contributions to the Hamiltonian, could be relatively important in estimating the accidental degeneracy of the two levels.
We use both our theoretical value and the recommended solar spectral value to illustrate the dependence of the average rate $\overline{A}_{MIT}$ on the magnetic field in Figure~\ref{Btr}. It is clear that even for relatively weak magnetic fields of only a few hundreds or thousands of Gauss, the $A_{MIT}$ will be significant compared to the competing M2-rate.
\subsection{Experimental Determination of The Energy Splitting}
\begin{figure}[t]
\centering
\includegraphics[width=0.8\textwidth]{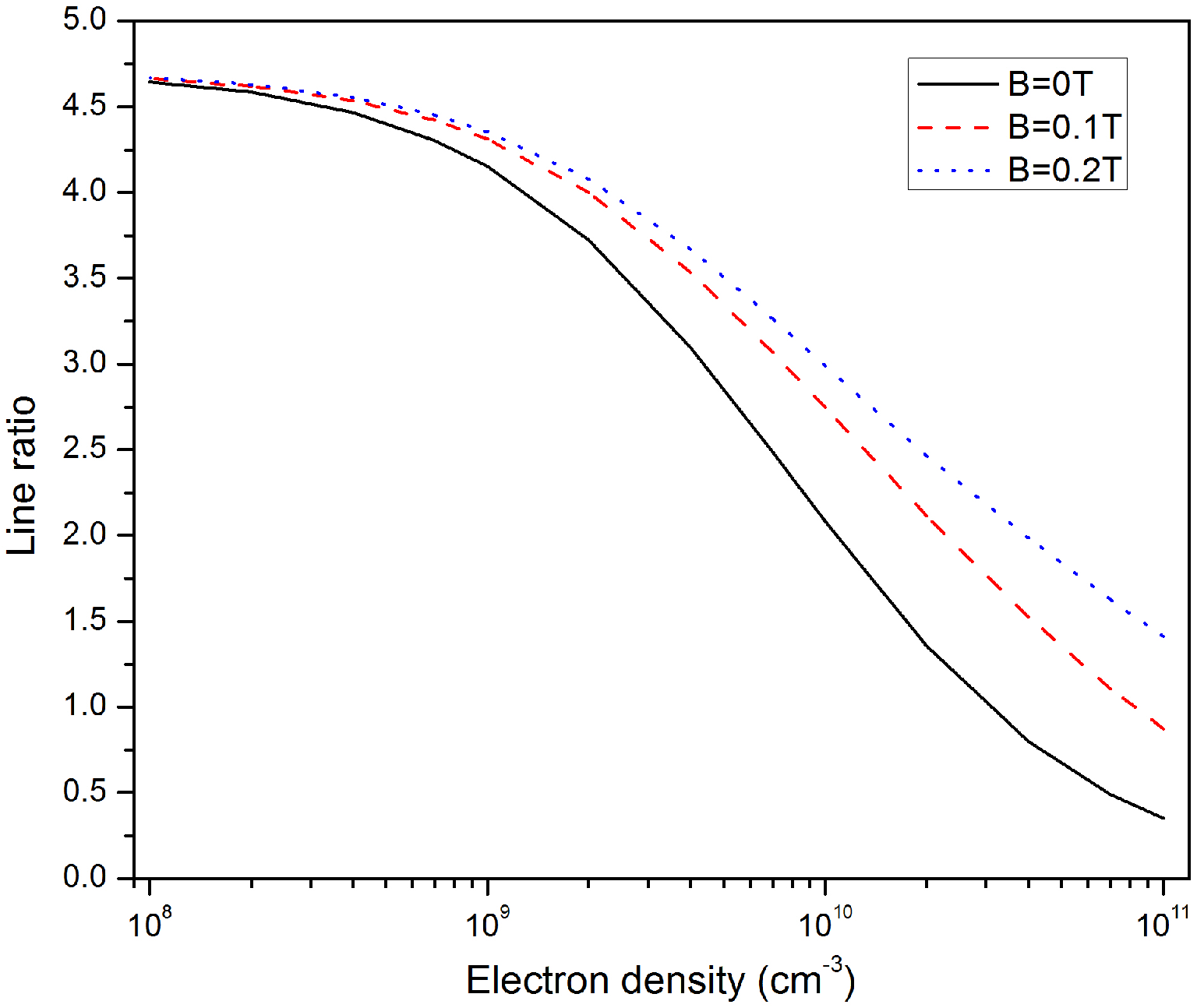}
\caption{\label{LnRD} The ratio of the rates for the magnetically induced $^4\mathrm{D}_{7/2}~\rightarrow~^2\mathrm{P}_{3/2}$ line and the allowed $^4\mathrm{D}_{5/2}~\rightarrow~^2\mathrm{P}_{3/2}$ line as a function of electron density and magnetic field in an EBIT. Calculations is for a monoenergetic beam energy of 250 eV and these data are displayed for some selected magnetic field strengths and over a range of densities ($10^8 - 10^{11} ~\mathrm{cm}^{-3}$) which covers the range for solar flares in the corona. Here we used the fine structure energy of 5 cm$^{-1}$.}
\end{figure}
To improve the accuracy of the estimated rate of the MIT, we need to turn to experiment for an accurate determination of the energy splitting. For this, we need to overcome two difficulties, first enough spectral resolution, and second a light-source with a low electron density and a magnetic field. The first requirement is not impossible to fulfill since the fine structure separation can be determined using a large spectrometer with a resolution of around 80¡¯000. This is far from the highest resolution achieved, since e.g. a spectrometer at the Observatory in Meudon has a resolution of 150¡¯000. However, the line from the $^4\mathrm{D}_{7/2}$ level has not been observed due to strict requirements on the light source. Most sources used at Meudon have generated too dense plasmas in which photon transitions from long lived levels cannot be seen (collisions are destroying the population of the upper state before the photon is emitted). In addition to this, observation of the $^4\mathrm{D}_{7/2} \rightarrow ^2\mathrm{P}_{3/2}$ line requires a strong enough magnetic field of, say a, tenth of a Tesla. This leads arguably to only two possible light sources on earth: Tokamaks ({Wesson 2004}) and the Electron Beam Ion Traps({Levine et al. 1988}) (EBITs). Tokamak plasma may be too dense, but since the magnetic fields involved are higher than what we are discussing here, the line might still be observable. However, the best choice for our purposes is an EBIT, which has an inherent magnetic field to compress the electron beam and is a low density light source. Although the Meudon spectrometer demonstrates that the required resolving power can be achieved this instrument is not compatable with the EBIT operating parameters and a dedicated instrument is required.

\begin{figure}[t]
\centering
\includegraphics[width=0.8\textwidth]{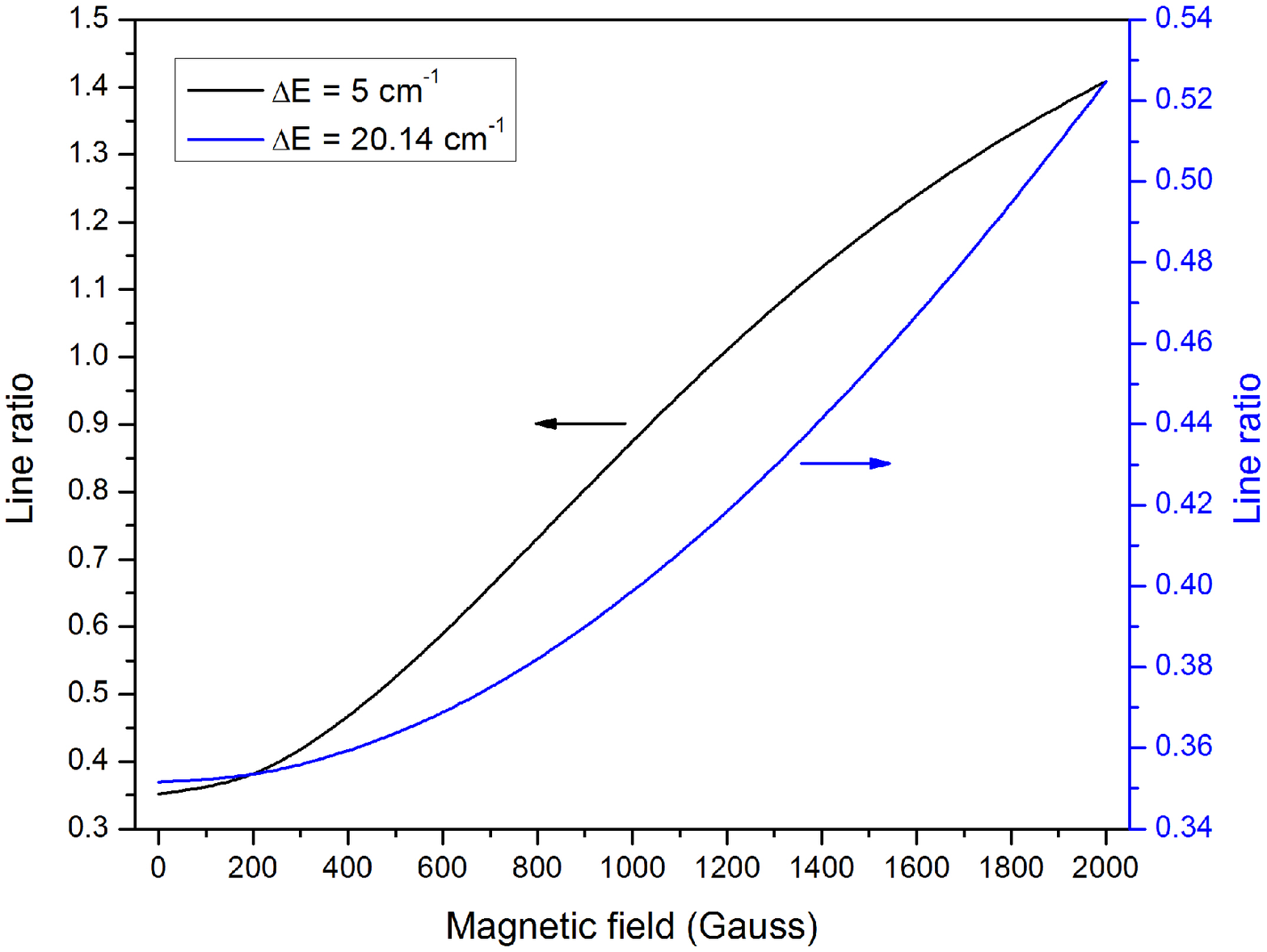}
\caption{\label{LnRB} Ratio of rates for the magnetically induced and the allowed transition as a function of magnetic field in an EBIT. The model is for a mono-energetic electron-beam energy of 250 eV and density of $1.0 \times 10^{11} ~\mathrm{cm}^{-3}$. We used the fine structure energy of 20.14 cm$^{-1}$ and 5 cm$^{-1}$. }
\end{figure}

To illustrate the usability of the EBIT source, we have made several model calculations to predict the relative strength of the two involved transitions under different circumstances. It should be made clear that the EBIT is a light source with a mono-energetic beam of electrons and that these models therefore are designed to predict conditions different from those in solar flares or the corona. It is important to remember that the intermediate goal before we can proceed is to propose an experiment to determine the crucial $^4\mathrm{D}_{7/2} - ^4\mathrm{D}_{5/2}$ energy separation. We present model calculations of the line ratio as a function of magnetic field and electron density (Figure~\ref{LnRD}) and magnetic field (Figure~\ref{LnRB}) of the EBIT, based on collisional-radiative modeling using the Flexible Atomic Code ({Gu 2008}). We show in Figure~\ref{LnRD}, for several magnetic fields, how the ratio of the rates of the magnetic-field induced $^4\mathrm{D}_{7/2} \rightarrow ^2\mathrm{P}_{3/2}$ and the allowed $^4\mathrm{D}_{5/2} \rightarrow ^2\mathrm{P}_{3/2}$ transitions varies as a function of the electron densities. It is clear that the magnetic-field induced line is predicted to be visible for the typical range of electron densities of an EBIT, that is $10^8 - 10^{11}$ $~\mathrm{cm}^{-3}$ (this happens to coincide with the range for solar flares). It is also clear from Figure~\ref{LnRB}, where we show the dependence of this ratio on the magnitude of the external magnetic field for a fixed density of $10^{11}$ $~\mathrm{cm}^{-3}$, that the line ratio will be sensitive to the magnetic field strength.

\section{Conclusion}
To conclude, in this paper we propose a novel and efficient tool to determine magnetic field strengths in solar flares. The method is useful for cases of low densities and small external magnetic fields (hundreds and thousands of Gauss) that have so far eluded determination. We illustrate that a spectral feature originating from the Fe$^{9+}$ ion is of special interest since it shows a strong dependence on the magnetic field strength, with two spectral lines drastically changing their relative intensities. We propose a laboratory measurement of the fine structure energy separation between the two involved excited states, a crucial parameter in the determination of the external field. When this energy separation has been established one can use our theoretical values for the reduced rate of the magnetic-field induced transition, which have an accuracy to within a few percent, to calculate the atomic response to the external magnetic field. Armed with this it is possible to design a space-based mission with a probe that could continuously observe and determine the reclusive magnetic fields of the solar flares.

\section{Acknowledgements}
This work was supported by the Chinese National Fusion Project for ITER No. 2015GB117000, Shanghai Leading Academic Discipline Project No. B107. We also gratefully acknowledge support from the Swedish Institute under the Visby-programme. WL and JG would like to especially thank the Nordic Centre at Fudan University for supporting their visits between Lund and Fudan Universities.
\section*{References}

\begin{thebibliography}{}

\bibitem[]{}
Aggarwal, K.~M., \& Keenan, F.~P. 2004,
\newblock {A\&A}, 427, 763.

\bibitem[]{}
Andersson, M., \& J\"{o}nsson, P. 2008,
\newblock {CoPhC}, 178(2), 156.

\bibitem[]{}
Andrew, K. L., Cowan, R. D., \& Giacchetti, A. 1967,
\newblock {JOSA}, 57(6), 715 .


\bibitem[]{}
Beiersdorfer, P., Scofield, J. H. \& Osterheld, A. L. 2003,
\newblock {PhRvL}, 90, 235003.

\bibitem[]{}
Bhatia, A., \& Doschek, G. 1995,
\newblock {ADNDT}, 60, 97.

\bibitem[]{}
Brosius, J., Davila, J., \& Thomas, R. 1998,
\newblock {APJS}, 119, 255.

\bibitem[]{}
Cheng, K. T., \& Childs, W. J. 1985,
\newblock {PhRvA}, 31(5), 2775.

\bibitem[]{}
Deb, N. C., Gupta, G. P., \& Msezane, A. Z. 2002,
\newblock {ApJS}, 141, 247.

\bibitem[]{}
Dong, C. Z., Fritzsche, S., Fricke, B., \& Sepp, W.-D. 1999,
\newblock {MNRAS}, 307, 809.

\bibitem[]{}
Grant, I. P. 2006, Relativistic Quantum Theory of Atoms and Molecules: Theory and Computation
\newblock {(Springer Series on Atomic, Optical, and Plasma Physics). Springer-Verlag New York, Inc., Secaucus,
NJ, USA.}

\bibitem[]{}
Grumer, J., Li, W., Bernhardt, D., et al. 2013,
\newblock {PhRvA}, 88, 022513.

\bibitem[]{}
Grumer J., Brage T., Andersson M., et al. 2014,
\newblock {PhyS}, accepted.

\bibitem[]{}
Gu M. F. 2008,
\newblock {CaJPh}, 86, 675.

\bibitem[]{}
Huang K.-N., Kim K., \& Cheng K.T. 1983,
\newblock {ADNDT}, 28, 355.

\bibitem[]{}
Ishikawa, Y., Santana, J. A., \& Trabert, E. 2010,
\newblock {JPhB}, 43, 074022.

\bibitem[]{}
J\"{o}nsson, P., Gaigalas, G., Biero, J., Fischer, C. F., \& Grant, I. 2013,
\newblock {CoPhC}, 184(9), 2197.

\bibitem[]{}
Levine, M. A., Marrs R. E., Henderson J. R., Knapp D. A. \& Schneider M. B., 1988,
\newblock {PhyS}, T22, 157-163.

\bibitem[]{}
Li, J., Brage T., J\"{o}nsson, P. \& Yang Y. 2014,
\newblock {arXiv:physics.atom-ph/1405.7787v1 (style for preprints before May 2014).}.

\bibitem[]{}
Li, J., Grumer, J., Li, W., et al. 2013,
\newblock {PhRvA}, 88, 013416.

\bibitem[]{}
Lin, H., Kuhn, J. R., \& Coulter, R. 2004,
\newblock {APL}, 613, L177.

\bibitem[]{}
Sandlin G.D. 1979,
\newblock {ApJ}, 227, L107.

\bibitem[]{}
Schrijver, C. J., DeRosa, M. L., Metcalf, T., et al. 2008,
\newblock {ApJ}, 675, 1637.

\bibitem[]{}
Smitt, R. 1977,
\newblock {SoPh}, 51, 113.

\bibitem[]{}
Stenflo, J. O. 1977,
\newblock {RPPh}, 41(6), 865.

\bibitem[]{}
Swings, P. 1943,
\newblock {ApJ}, 98, 116-128.

\bibitem[]{}
Thomas, R., \& Neupert, W. 1994,
\newblock {APJS}, 91, 461.

\bibitem[]{}
Wesson, J., 2004,
\newblock { Tokamaks, (Oxford University Press.).}

\bibitem[]{}
White, S. M. 2004, Coronal Magnetic Field Measurements Through Gyroresonance Emission, Solar and Space Weather Radiophysics, (D. Gary, C. U. Keller Editors, Astrophys. And Space Science Library.).

\bibitem[]{}
White, S. M., \& Kundu, M. R. 1997,
\newblock {SoPh}, 174, 31-52.

\bibitem[]{}
Wood, D. R., Andrew, K. L., Giacchetti, A., \& Cowan, R. D. 1968,
\newblock {JOSA}, 58(6),830.

\bibitem[]{}
Del Zanna, G., Berrington, K. A. \& Mason H. E. 2004,
\newblock {A\&A}, 422, 731.

\bibitem[]{}
Del Zanna, G., Storey, P. J., Badnell, N. R., \& Mason, H. E. 2012,
\newblock {A\&A}, 541, A90.

\end{thebibliography}

\end{document}